\documentclass[12pt,epsf]{article}
\usepackage{graphicx}
\begin{document}

%%%%%%%%%%%%%%%%%%%%%%%%%%%%%%%%%%%%%%%%%%%%%%%%%%%%%%%%%%%%%%%%%%%%
%%%%%%%%%%%%%%%%%%%%%%%%%%%%%%  Defs. %%%%%%%%%%%%%%%%%%%%%%%%%%%%%%
%%%%%%%%%%%%%%%%%%%%%%%%%%%%%%%%%%%%%%%%%%%%%%%%%%%%%%%%%%%%%%%%%%%%
\def\a{\alpha}
\def\b{\beta}
\def\c{\varepsilon}
\def\d{\delta}
\def\e{\epsilon}
\def\f{\phi}
\def\g{\gamma}
\def\h{\theta}
\def\k{\kappa}
\def\l{\lambda}
\def\m{\mu}
\def\n{\nu}
\def\p{\psi}
\def\q{\partial}
\def\r{\rho}
\def\s{\sigma}
\def\t{\tau}
\def\u{\upsilon}
\def\v{\varphi}
\def\w{\omega}
\def\x{\xi}
\def\y{\eta}
\def\z{\zeta}
\def\D{\Delta}
\def\G{\Gamma}
\def\H{\Theta}
\def\L{\Lambda}
\def\F{\Phi}
\def\P{\Psi}
\def\S{\Sigma}

\def\o{\over}
\def\beq{\begin{eqnarray}}
\def\eeq{\end{eqnarray}}
\newcommand{\gsim}{ \mathop{}_{\textstyle \sim}^{\textstyle >} }
\newcommand{\lsim}{ \mathop{}_{\textstyle \sim}^{\textstyle <} }
\newcommand{\vev}[1]{ \left\langle {#1} \right\rangle }
\newcommand{\bra}[1]{ \langle {#1} | }
\newcommand{\ket}[1]{ | {#1} \rangle }
\newcommand{\EV}{ {\rm eV} }
\newcommand{\KEV}{ {\rm keV} }
\newcommand{\MEV}{ {\rm MeV} }
\newcommand{\GEV}{ {\rm GeV} }
\newcommand{\TEV}{ {\rm TeV} }
\def\diag{\mathop{\rm diag}\nolimits}
\def\Spin{\mathop{\rm Spin}}
\def\SO{\mathop{\rm SO}}
\def\O{\mathop{\rm O}}
\def\SU{\mathop{\rm SU}}
\def\U{\mathop{\rm U}}
\def\Sp{\mathop{\rm Sp}}
\def\SL{\mathop{\rm SL}}
\def\tr{\mathop{\rm tr}}

\def\IJMP{Int.~J.~Mod.~Phys. }
\def\MPL{Mod.~Phys.~Lett. }
\def\NP{Nucl.~Phys. }
\def\PL{Phys.~Lett. }
\def\PR{Phys.~Rev. }
\def\PRL{Phys.~Rev.~Lett. }
\def\PTP{Prog.~Theor.~Phys. }
\def\ZP{Z.~Phys. }

\def\stau{\widetilde{\tau}}

%%%%%%%%%%%%%%%%%%%%%%%%%%%%%%%%%%%%%%%%%%%%%%%%%%%%%%%%%%%%%%%%%%%%

\baselineskip 0.7cm

\begin{titlepage}

\begin{flushright}
UT-06-15 \\
TKYNT-06-15
\end{flushright}

\vskip 1.35cm
\begin{center}
{\large \bf
Stau-catalyzed Nuclear Fusion
}
\vskip 1.2cm
K. Hamaguchi${}^{1}$, T. Hatsuda${}^{1}$ and T. T. Yanagida${}^{1,2}$
\vskip 0.4cm

${}^1${\it Department of Physics, University of Tokyo,\\
     Tokyo 113-0033, Japan}

${}^2${\it Research Center for the Early Universe, University of Tokyo,\\
     Tokyo 113-0033, Japan}

\vskip 1.5cm

\abstract{
We point out that the stau $\stau$ may play a role of a catalyst 
for nuclear fusions
if the stau is a long-lived particle as in the scenario of 
gravitino dark matter. In this letter,
we consider $dd$ fusion under the influence of $\stau$
  where the fusion is enhanced because of a short distance between the two
deuterons. We find that one chain of the $dd$ fusion may release an energy 
of $O(10)$ GeV per stau. We discuss problems of making the
 $\stau$-catalyzed
nuclear fusion of practical use with the present technology of producing stau.
 }

\end{center}
\end{titlepage}

\setcounter{page}{2}

\section{Introduction}

Supersymmetry (SUSY) is the most attractive candidate beyond the standard 
model (SM) \cite{SUSY}. It provides not only a natural solution to 
the hierarchy problem in the SM, 
but also it may explain the dark matter (DM) density in the 
universe.
Among several candidates for the DM, the gravitino is very attractive, 
since the presence of the gravitino is an inevitable prediction of 
the supergravity. Furthermore, if the gravitino is indeed the stable
lightest SUSY particle, we may avoid a serious cosmological problem, 
so-called the gravitino problem \cite{gravitino-BBN}. 

In the case that the gravitino is the stable DM in the universe, 
the next lightest SUSY particle (NLSP) has a long lifetime, since 
it decays to gravitino through very weak interactions suppressed by the Planck scale.
The most attractive candidate of NLSP is the scalar partner
of the tau lepton called stau, $\stau$. It
 provides us with a test of 
the supergarvity in future collider experiments such as LHC \cite{BHRY}.
The lifetime of the stau is determined by the masses of stau and gravitino
as
 \begin{eqnarray}
 t_{\stau} \simeq 0.2~{\rm years}
 \left({m_{3/2}\over 10~{\rm GeV}}\right)^2
 \left(100~{\rm GeV}\over {m_{\stau}}\right)^5
 \left({1-{m_{3/2}^2\over m_{\stau}^2}}\right)^{-4}.
\end{eqnarray} 
For instance, the lifetime is about $0.9$ years
 for $m_{\stau} \simeq 100$ GeV and $m_{3/2}\simeq 20$ GeV.\footnote{
 The decay of such a long-lived particle may spoil the success of the standard cosmology.
Furthermore, it has been recently pointed out that the
staus can form bound states 
with light elements and may affect the nuclear reaction rate in 
 the big-bang nucleosynthesis~\cite{CBBN}. 
 However, these effects can be avoided if there is 
a late-time entropy production to dilute the stau relic 
abundance~\cite{Buchmuller:2006tt}. }

 In this short letter, we point out that the long-lived stau
 (in general, a  long-lived and negatively charged heavy particle)
 can be used as a catalyst for nuclear fusion such as 
 $d + d\rightarrow$ $^3{\rm He} + n$, $t +p$, $^4{\rm He} + \gamma$. 
 A crucial point here is that the negative $\stau$ captures a $d$ to form a neutral atomic system
 and the Coulomb repulsion force between the bounded $d$ and a free $d$ is screened. 
 Thus, the free $d$ can reach close to the bounded $d$ and the distance between
 them can be as short as $\sim 50~{\rm fm}$ or less.
 Because of the short distance between two $d$'s in this system, the nuclear fusion is 
 substantially enhanced \cite{Zweig, Okun}. 
 After the fusion taking place,  the produced nuclei ($A = p,~t,~^3{\rm He}$ and $^4{\rm He}$)
 with the momentum of about 24 - 75 MeV escape from the
 Coulomb potential of stau and hence we can reuse the stau again as a catalyst
 to start another cycle inside the liquid or solid deuterium. In reality, 
 the fusion product $A$ has a small probability of  $O(10^{-4})$
 to be trapped by stau as discussed in section 3. This terminates 
the fusion chain after $O(10^4)$ cycles. Since  one fusion releases 3 - 4 MeV, 
 one can produce energy of $O(10)$ GeV per stau.
 This may be compared with a similar case of the muon($\mu$)-catalyzed
 fusion with the use of the molecular state such as
 $\mu dt$ where only about $10^2$ cycles (a few GeV energy release) 
 are possible per muon \cite{MCF}.

 Unfortunately, the required energy to produce one stau 
 by high-energy muon-nucleon scattering is extremely larger than the above energy gain 
 as discussed in the last section. Even assuming that one can 
 develop a device to reactivate stau
 by stripping off the trapped nuclei $A$ and to continue the fusion cycle,
 it turns out that the life-time of the stau should be longer than 300 years
 in our estimate to make the $\stau$-catalyzed fusion of practical use. 
 This is because the time scale of one cycle, which is determined
 by the reaction time of $\stau d + d$ system, is about $2~\mu$sec.
 Thus, the present $\stau$-catalyzed fusion does not become a promising 
 source of energy in this century, unless we find a more efficient
 mechanism/technology for the stau production.
 We briefly comment on such a  possible mechanism using a high-energy
 neutrino beam. 

\section{$\stau d$ atom and $\stau d$-$d$ interaction }

 For later purpose, let us consider the heavy analogue of the H atom (hydrogen) and
  the ${\rm H}^-$ atom (a system made of a proton and two electrons)
  and summarize their properties.
In our case, the stau corresponds to the proton and $d$ to the electron. 
The only differences are the masses and spins of the constituent particles:
 the stau and $d$ have spin 0 and 1, respectively.
The spins play no crucial role on forming non-relativistic bound states.
Concerning the masses,  it has been proven that,
in non-relativistic approximation with only the Coulomb interaction,
a three-body system with charges 
$q_i=\pm 1$, $\mp 1$, $\mp 1$ and masses $M$, $m$, $m$ has at least one bound state for
 arbitrary ratios of $m$ and $M$, and has only one bound state for 
 $m/M \lsim 0.2101$~\cite{RNHill}.
 Therefore, the $\stau d d$ system has only one  bound state because of 
 $m/M\simeq 0.02~(100~\mathrm{GeV}/m_{\stau})$.

In Table.~\ref{tab:staudd-pee},  $\stau d$ and 
 $\stau d d$ are compared to  H($pe$) and H$^-$($pee$), respectively.
Note that the size of the $\stau d$ atom is an order of magnitude 
smaller than the size of the muonic atom $\mu d$, 
and hence we may expect an enhanced nuclear fusion rate.

\begin{table}[t]
\begin{tabular}{|cccc|ccc|}
\hline
System & $\mu_{r}$ & $a_B$ & $E_b^{(2)}$ 
& System & $R$ & $E_b^{(3)}-E_b^{(2)}$
\\ \hline \hline
${\rm H}(pe)$ & $0.51$~MeV & $5.3\times 10^4$~fm & $14$~eV
& ${\rm H}^-(pee)$ & $\sim 16\times 10^4$~fm & $0.75$~eV
\\
$\stau d$ & $1.9$~GeV & $15$~fm & $50$ keV
& $\stau d d$ & $\sim 45$~fm & $2.8$~keV
\\ \hline
\end{tabular}
\caption{Comparison of $\stau d$ to H($pe$) and $\stau d d$ to H$^-$($pee$). $\mu_r$, 
$a_B$, and $E_b^{(2)}$ are the reduced mass, Bohr radius, and binding energy of the two--body system, respectively.
$R$ is the typical size of the three--body system, which is about $\sim 3 a_B$~\cite{Lin}. 
$E_b^{(3)}-E_b^{(2)}$ is the extra binding energy of the three--body system with respect to the
 two--body system, which is about $0.055\times E_b^{(2)}$ for small mass ratio $m/M\ll 1$ 
 (cf. \cite{Lin}). 
}
\label{tab:staudd-pee}
\end{table}

\begin{figure}[bhtp]
  \begin{center}
    \includegraphics[keepaspectratio=true,height=60mm]{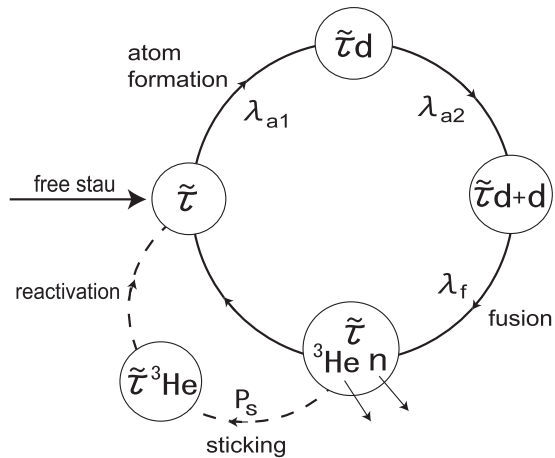}
  \end{center}
  \caption{An example of the $\stau$-catalyzed fusion cycle.  
  $\lambda_{a1}$, $\lambda_{a2}$ and $\lambda_f$ are the 
  formation rate of $\stau d$ atom, the interaction rate of the $\stau d$+$d$
   system, and the reaction rate of the $dd$ fusion. $P_s$ is the 
    sticking probability of $^3$He trapped by $\stau$.  }
  \label{fig:fusion_cycle}
\end{figure}

  To show how the $\stau$-catalyzed fusion proceeds, a typical fusion cycle 
  with $dd\rightarrow$$^3{\rm He} + n$ reaction is schematically shown
  in Fig. \ref{fig:fusion_cycle}.
  Suppose that a free $\stau$ is stopped inside the cold deuterium 
  with the typical density of the liquid hydrogen (4.25 $\times 10^{22}$ atoms/cm$^3$).
  Then the formation of the 1s state of the $\stau d$ atom occurs quickly with the time
  scale of 1 ps ($=10^{-12} $ sec) through the capture process to the higher orbit,
  $\stau + (de^-) \rightarrow (\stau d) + e^-$, followed by the de-excitation to the 1s state.
  The capture time estimated by the geometrical size of $d e^-$ leads to 0.1 ps 
  which is comparable to the formation of $\mu d$ atom in the 
  same environment.  The subsequent de-excitation time scale, if the standard dipole 
  radiation formula is used, is of order 1 ps, which is about 10 times faster than that of 
  $\mu d$ \cite{MCF} mainly because of the large binding energy of $\stau d$.
  Thus the formation rate, $\lambda_{a1}$, in Fig.\ref{fig:fusion_cycle} is estimated
  to be about 10$^{12}$ sec$^{-1}$.

  On the other hand, the typical rate for $\stau d$ to interact with $d$
  (as shown by  $\lambda_{a2}$ in Fig.\ref{fig:fusion_cycle}) 
  is much lower because $\stau d$ is charge neutral and of small size.
  Since the typical velocity of the liquid deuterium\footnote{For
   temperature of order $T=20$K, the typical velocity of the 
  deuteron is   $v \simeq \sqrt{3kT/m_d} \simeq 1.7 \times 10^{-6}$.} 
 is much smaller than the typical velocity of $d$ inside the $\stau d$ atom, 
  the structure of $\stau d$ would be substantially modified when $d$ approaches 
  adiabatically to $\stau d$. Accordingly, the incoming $d$ may start to 
  feel attraction not at the distance of about $a_B(d)=15$ fm (the Bohr radius of $\stau d$)
  but rather at the distance of  about the 
   size of the $\stau d d$ atom $\sim  3 \times a_B(d)$. 
   (This
    should be, however, confirmed by the exact three-body calculation of the 
   $\stau d -d$ scattering.)  
  By adopting $R \equiv 3a_B(d) = 45$ fm as an ``effective" interaction range,
   we find the low energy cross section between $\stau d$ and $d$ as
    $\sigma_{\stau d +d} \simeq 4 \pi R^2 \simeq 2.5 \times 10^{-22}~{\rm cm}^2$. 
    With the  density
    $n\simeq 4.25\times 10^{22}$ atoms/cm$^3$, this leads to
    $\lambda_{a2} \simeq n \sigma v\simeq  5.5 \times 10^5~{\rm sec}^{-1}$.\footnote{The
     formation of the $\stau d d$ atom in $\stau d$+$d$ reaction requires a 
    photon emission and has $O(10^{-5})$ times smaller cross section than 
    $\sigma_{\stau d +d}$, 
    which is obtained 
    by rescaling the formulae in Refs.~\cite{Massey,Chandrasekhar}.
  Since the $dd$ fusion rate is much larger than
  $\lambda_{a2} \sim 5.5 \times 10^5~{\rm sec}^{-1}$, 
  the fusion takes place before the $\stau d d$ atom is formed.}

\section{Nuclear fusion and sticking}

In this section we consider three major reactions:
  \begin{eqnarray}
 {\rm (i)}:\ \ \  d+d  &\rightarrow&  ^3{\rm He} + n \ \ \ \  (Q=3.3 \MEV), \nonumber \\
 {\rm (ii)}:\ \ \  d+d &\rightarrow&   t + p \ \    \ \ \ \ \ \ (Q=4  \MEV),\nonumber \\
 {\rm (iii)}:\ \ \ d+d &\rightarrow&  ^4{\rm He} + \gamma \ \ \ \ (Q=24 \MEV), \nonumber 
  \end{eqnarray}

 The elementary cross section $\sigma(E)$ as a function of the 
 center-of-mass kinetic energy $E$  of the above reactions 
 can be written as $ \sigma(E) = \frac{S(E)}{E} \ e^{-2\pi \eta(E)}$.
Here $e^{-2\pi \eta}$ is the Gamow tunneling factor of the 
 Coulomb barrier  with the Sommerfeld parameter,
  $\eta(E) = \alpha Z_1 Z_2 /  v = \alpha Z_1 Z_2 \sqrt{\mu_{r}/(2E)}$.
 $Z_1$ and $Z_2$ are the charge numbers of incoming particles ($Z_1=Z_2=Z_d=1$
 in the present case), $v$  the relative velocity and $\mu_r$ 
  the reduced mass $\mu_{r} = \frac{m_1 m_2}{m_1+m_2}$ 
  ($\mu_{r}=m_d/2$ in the present case).
 $S(E)$ is the spectroscopic factor which can be
 extracted  from the analyses of the low-energy nuclear reaction data as
 $S(1~ \KEV < E < 50~\KEV) \simeq 50~\KEV \cdot {\rm barn}$ for
 (i) and (ii), and  $S(10~\KEV < E < 100~\KEV) \simeq 5 \times 10^{-6}~\KEV \cdot {\rm barn}$ for (iii) \cite{NACRE}.
The typical fusion reaction rate $\lambda_f$ in the $\stau d$ + $d$ system
  may be roughly estimated as follows:
   $ \lambda_f \sim \sigma(\bar{E} ) \times \bar{v}   \times 1/V$, 
  where  $\bar{E}$ and $\bar{v}$ are the typical
   center-of-mass kinetic energy and relative velocity of the deuterons, respectively.
   $\bar{v}/V$ is a flux factor, so that $V$ is a typical volume in which
    the fusion takes place ($V \sim \frac{4\pi}{3} R^3  \ {\rm fm}^3$ with
     $R = 45 {\rm fm}$).
 $\bar{E} = 2 \bar{E}_d \sim (1/m_d R^2) \sim 10  \KEV$ from uncertainty principle.
 Similarly, we have $\bar{v} =  \sqrt{2 \bar{E}/\mu} = \sqrt{4 \bar{E}/m_d}
 \sim   5 \times 10^{-3}$.
 Therefore we obtain
  $\lambda_{f{\rm (i,ii)}} \sim  10^{14} \ {\rm sec}^{-1}$ and  
 $\lambda_{f{\rm (iii)}} \sim 10^7 \ {\rm sec}^{-1}$, which are
  much larger than $\lambda_{a2}$ in the previous section. Note that 
  $\lambda_{f{\rm (i,ii,iii)}}$ increases by factor $10^{2-3}$ if
   we use $R=a_B(d)=15$ fm instead of 45 fm.

 After the fusion reactions (i)-(iii),
 positively charged products ($A= p, t$,$^3{\rm He}$,$^4{\rm He}$) 
  may be trapped by $\stau$ and form a Coulomb bound state $\stau A$.
   If this ``sticking" process happens, 
   the fusion chain  will be terminated. 
 The probability of the sticking,  $P_s$, gives a 
  stringent constraint on  the number of fusion reactions per $\stau$.
  For the muon-catalyzed fusion,
   $P_s$ is known to be about $15 \ (1)\  \%$ for $\mu dd$ ($\mu dt$) \cite{Jackson,MCF}.
 In the sudden approximation where the instantaneous $dd$ fusion 
 does not affect the state of $\stau$,  the sticking probability 
 for infinitely heavy $\stau$ may be estimated from
  an overlap integral of the initial and final state wave functions of $A$ \cite{Migdal}:
\begin{eqnarray}
 P_s = \left| \int \Psi_{\rm 1s}^* ({\bf r}) \Phi_{k} ({\bf r}) d^3 r \right|^2.
\label{eq:sticking-P}
\end{eqnarray}

  Here  $\Psi_{\rm 1s}({\bf r}) \propto \exp(-r/a_0)$ is the final state
  1s wave function of  the $\stau A$ atom. The Bohr radius
   $a_{0}$ reads approximately $2a_B(d), 2a_B(d)/3, a_B(d)/3$
    and $a_B(d)/4$ for $A= p, t$, $^3{\rm He}$ and $^4{\rm He}$, respectively.
  ($a_B(d)$ denotes the Bohr radius of the $\stau d$ atom.)
   We parametrize the wave function 
   of a nucleus $A$ just formed, $\Phi_{k} ({\bf r})$, as a product of the
   plane wave with momentum $k$ and a wave packet of a size $R_A$; $\Phi_{k}
  ({\bf r}) \propto \exp (i {\bf k}\cdot {\bf r}) \times  \exp (-r/R_A)$  \cite{Zweig}.

  The momentum $k$ is obtained from 
 the $Q$-value of each reaction:  $k \simeq 68$ MeV, 75 MeV 
  and 24 MeV, for (i), (ii), and (iii). 
  The size of the wave packet $R_A$ over which the nucleus $A$ is distributed 
  is identified with the effective interaction range $R=3a_B(d)=45$ fm instead of 
  the Bohr radius  of $(\stau d)$ according to the discussion in Section 2.
  
  With all the above inputs,
  the integral in Eq.(\ref{eq:sticking-P}) can be carried out
  analytically and the sticking probabilities for 
  $\stau p, \stau t$, $\stau ^3{\rm He}$ and $\stau ^4{\rm He}$
  read $P_s = 2\times 10^{-7}, 2 \times 10^{-5}, 4 \times 10^{-4}$ and $10^{-2}$,
   respectively. By multiplying these numbers to the fusion rate, 
  the sticking of $t$ and $^3{\rm He}$ gives 
  the largest rate to terminate the fusion chain (cf. Fig.~\ref{fig:fusion_cycle}).
  The small sticking probabilities obtained here (e.g. $10^{-2}$ smaller than that given
   in \cite{Zweig} for $^3 {\rm He}$ case)
   are primarily because $\stau d$ and $d$ start to interact
    much outside the size of $(\stau d)_{1s}$ (i.e. $ R= 3a_B(d) > a_B(d)$),
   and hence $A$ is formed rather  far away from $\stau$ (i.e. $R_A \sim R > a_B(d)$).
   
So far, we have not considered the fact
  that flux of the slow $d$ coming from outside of the $\stau d$ atom
   is reduced due to the nuclear $dd$ fusion. Namely $d$ may not 
   penetrate deep inside the $\stau d $ system.
  This may introduce an extra suppression factor $\kappa$
   to $P_s$. Also this effect was not considered in previous works
  \cite{Zweig,Okun}. Magnitude of $\kappa$ is unknown at present and should be
   derived from the exact three body calculation of $\stau d -d$ system
    with appropriate nuclear fusion process.\footnote{Different
   way of evaluating the fusion cross section
   and sticking probabilities is given in \cite{Sawicki}. Both the 
   effective range of the interaction $R$ and the $\kappa$-factor
    are not take into account in this work.} 
 Taking into account all  the factors, the energy
  production $E_{\stau dd}$ per $\stau$ is estimated as  
\begin{eqnarray}
 E_{\stau d + d} \sim \frac{ \frac{1}{2}(3.3 + 4)  \ {\rm MeV}}
 {\frac{1}{2}(4 \times 10^{-4} + 2 \times 10^{-5}) \times \kappa}
 \simeq  20\ {\rm GeV}/\kappa.
 \end{eqnarray}

\section{Discussion}     

Let us now discuss a possible production of the staus in the laboratory.
 We consider the  $\mu + N \ ({\rm nucleon})$  scattering with a fixed nuclear target.
 The stau-production cross section depends on the spectrum of 
 SUSY particles. Here, we adopt an optimistic situation discussed
 in \cite{production} where the slepton-production cross section
is $O(1)$~pb for the laboratory energy of the muon $\mu$,  $E_\mu \simeq 1000$ TeV. 
Since all SUSY particles decay quickly to the staus, the 
 stau-production cross section is also of 
 $O(1)$~pb.  

 Assuming Fe target of 200 m length with the nucleon density $n_N\simeq 5\times 10^{24}/{\rm cm}^3$,
 the number of produced staus per muon of $E_\mu\simeq 1000$ TeV reads
\begin{eqnarray}
 n_{\stau} \simeq \sigma \times n_N\times 200~{\rm m}\simeq 10^{-7}.
 \label{eq:nstau}
\end{eqnarray}
Notice that the stopping range of the muon
inside the Fe target is $O(1)$ km for $E_\mu\simeq 1000$ TeV~\cite{Eidelman:2004wy,Ahlers:2006pf}.
Eq.~(\ref{eq:nstau}) implies that we need $10^7 \times 1000$ TeV to produce
a single stau. On the other hand, one stau could produce $O(10)$ GeV
energy for a single chain of the $dd$ fusion 
as we have discussed in section 3.
 Therefore, to make the present stau-catalyzed
 $dd$ fusion to be of practical use, we need to recycle the stau
  at least 10$^{12}$ times. For this purpose, we 
 should collect the inactive $\stau A$ atoms and strip the nuclei $A$ 
 from the staus. It is beyond the scope of the present short
 letter to investigate possible reactivation mechanisms.

 As discussed in section 2, the time scale of the one cycle of the stau-catalyzed
 $dd$ fusion is dominated by the time scale of $\stau d+d$ interaction which is about
  2 $\mu$s. 
 This leads to the time scale of a single chain as 
 2 $\mu$sec/($2 \times 10^{-4} \times \kappa$) 
 which is  larger than 0.01 sec. 
  Assuming that we find a sufficiently fast reactivation mechanism of stau,
  the life time of the stau should be longer than 300 years at least for 
  the output energy to exceed the input energy.
  Notice that this estimate is independent of $\kappa$.
 In order to  make the present $\stau$-catalyzed nuclear fusion an interesting source of energy,
 we thus need to find a more efficient
 mechanism and/or technology for the stau production.

Here we speculate on a possible mechanism of producing staus more efficiently. 
That is, we consider the  
  muon-neutrino beam (e.g. from the muon storage ring) injected into the earth as a target.
The mean free path of the neutrino is about the
diameter of the earth ($\sim 10^4$ km) provided that the
weak-interaction cross section is $1000$ pb at $E_\nu\simeq 1000$ TeV,
and one can produce $3\times 10^{-3}$ staus per neutrino for SUSY cross section $1$
pb~\cite{Ahlers:2006pf}. 
 We may  capture the produced staus because the size of the neutrino beam on the other side
of the earth is only 1 m~\cite{Sugawara:2003wa}. Taking into account 
the fact that the mean free path of the stau
 inside the earth is about 10$^3$ km, we may obtain $3\times 10^{-4}$ staus per neutrino.
  Namely, the  number of staus is $3 \times 10^3$ times larger than that produced by the muon beam, and hence
   the life-time of the stau can be less than a year to produce net energy 
   gain.\footnote{This is, of course, an optimistic estimate. In reality, it will not be easy to capture all
   the high-energy staus produced in the earth.
    Also, it will cost more than 1000 TeV to create the 1000 TeV neutrino.
   Moreover, constructing the 1000 TeV neutrino factory is certainly beyond the current technology.}

 To make the $\stau$-catalyzed fusion as a real energy source,    
 we certainly need  to develop revolutionary technology 
  of creating many staus with low cost and of reactivating $\stau A$ atoms in a fast and 
  efficient way.   Despite all these difficulties, 
  the $\stau d d$ fusion cycle proposed in this letter is itself
  an interesting interdisciplinary phenomenon which is worth to be studied further.
  Other than catalyzing the nuclear $dd$ fusion, negatively charged $\stau$ may 
  also provide a new tool in nuclear physics.
  Indeed, if the stau $\stau$ is embedded in heavy nuclei, 
   it will form exotic Coulomb bound states with their level structures
    affected by the charge distribution of the nuclear interior. 
  Namely, the stau may be used as a probe of the deep interior of heavy nuclei.

\section*{Acknowledgment}
We are grateful to A.~Ibarra  and M. Kamimura for helpful comments and
 suggestions.
  T.H. was partially supported by JSPS grant No.18540253.

\end{document}